\documentclass[a4paper,12pt]{article}
\usepackage[utf8]{inputenc}
\usepackage{amssymb,amsmath,latexsym}
\usepackage{graphicx}
\usepackage{psfrag}
\usepackage{hyperref}
\usepackage{mathtext}
\usepackage{longtable}
\usepackage[top=2cm, bottom=2cm, left=3cm, right=1.5cm]{geometry}
\begin{document}
\title{Cosmological dynamics of fourth order gravity with a Gauss-Bonnet term}
\author{Mikhail M. Ivanov$^{1,2}$\footnote{E-mail: mm.ivanov@physics.msu.ru}, Alexey V. Toporensky$^2$\footnote{E-mail: atopor@rambler.ru}}
\date{}
\maketitle
\begin{center}
$^1$ {\it Physical Faculty, Moscow State University, Moscow, 119991, Russia}\\
$^2$ {\it Sternberg Astronomical Institute, Moscow State University, Universitetsky prospect, 13, Moscow, Russia}
\end{center}
\begin{abstract}
We consider cosmological dynamics in fourth order gravity with both  $f(R)$ and  $\Phi(\mathcal G)$ correction to the
Einstein gravity ($\mathcal{G}$ is the  Gauss-Bonnet term). The particular case for  which both terms are equally important on power-law solutions is described. These solutions and their stability are studied using the dynamical system approach. We also discuss the
condition of existence and stability of a de Sitter solution in a more general situation of power-law $f$ and $\Phi$. 
\end{abstract}
\section{Introduction}
Theories of modified gravity become very popular in the beginning of our century
in attempts to explain the accelerated expansion of our Universe which is proved observationally
\cite{Obs1, Obs2}. Another motivation is incorporating quantum corrections to General
relativity (GR) which are believed to be present in the form of terms of higher order in the curvature.

It is known that introducing additional terms, proportional to the Kretchmann invariant
$R^{\mu\nu\sigma\rho}R_{\mu\nu\sigma\rho}$ into the action often leads to solution with increasing spatial
anisotropy \cite{BH, TT} incompatible with the known picture of our Universe.
This is the reason why modified gravity theories often deal with only $f(R)$-corrections
to Hilbert-Einstein action. However, special properties of the Gauss-Bonnet combination
$\mathcal{G}=R^{\mu\nu\sigma\rho}R_{\mu\nu\sigma\rho}-4R^{\mu\nu}R_{\mu\nu}+R^2$ make $f(\mathcal{G})$ correction terms reasonable to study.
Since $\mathcal{G}$ is a topological invariant in 4 dimensions, terms linear in $\mathcal{G}$ do not modify
the equations of motion (if $\mathcal{G}$ enters into the action in a product with function of another dynamical
variable, like in string-gravity corrections, it modifies equations of motion and gives rise to interesting effects \cite{Kawai, Maeda, AS}). Cosmological effects of $f(R)$ correction terms have been
studied recently for a number of different functions $f$. 

In the present paper we consider
both $f(R)$ and $\Phi(\mathcal{G})$ corrections to the Einstein gravity, find a situation where
they are equally important and outline differences of cosmological behaviour in such combined
model from models with only $f(R)$ or $\Phi(\mathcal{G})$ terms in the action. 

Viable cosmological models have been constructed for each theory separately (see, e.g., \cite{ON1,CENOZ,Barrow,dFS} for $\Phi(\mathcal{G})$). Furthermore, finite-time singularities (\cite{BO}), cosmological perturbations (\cite{dFGS}), stability around a spherically symmetric static space-time (\cite{dFST}),  and the number of  degrees of freedom and ghost problems (\cite{dFT}) have already been  investigated. Recently more general theories with $\mathcal{F}(R,\mathcal{G})$ have become a subject of investigation
\cite{dFST,Ali}. 
The review \cite{ON2} considers some more complicated versions of modified gravity including non-local theories and solutions which can be of interest from the viewpoint of cosmic acceleration have been described.
Our goal is different: we try to consider the general picture of cosmological dynamics in modified gravity theories applying the Dinamical system approach, used actively in  $f(R)$ gravity \cite{AGPT, CDCT, CTD}.
We make the next step taking into account the Gauss-Bonnet term restricting ourself in the present paper to a particular case where the function $\mathcal{F}(R, \mathcal{G})$ is a sum of a function of $R$ and a function of $\mathcal{G}$.
We find asymptotic regimes and studying their stability in such a modified Gauss-Bonnet gravity theory.

\section{Basic equations}
The action we study is given by\footnote[1]{The signature of the metric is assumed to be $(-,+,+,+)$ and the speed of light $c$ is taken to be equal to 1.}
	\begin{equation}
		\label{eq:1}
			\int \frac {  f(R) + \Phi(\mathcal{G}) }  {16 \pi G_N}\sqrt{-g} \, d^4x + S_{M},
			\end{equation}
where $f(R)$ and $\Phi(\mathcal{G})$ are general differentiable functions of the Ricci scalar $R$ and the Gauss-Bonnet invariant $\mathcal{G}$, $G_N$ is the Newton constant. Variation of this action with respect to $g^{\mu \nu}$ leads to the following field equations:
	\begin{equation}
		\label{eq:2}
			R_{\mu\nu} - \frac{1}{2}Rg_{\mu\nu} - \Sigma_{\mu\nu} = 8\pi G_NT_{\mu\nu} \,,
			\end{equation}
where
	\begin{equation}\begin{split}
		\Sigma_{\mu\nu} = & \nabla_\mu\nabla_\nu F-g_{\mu\nu} \square F -2g_{\mu\nu} R \square\xi +2R\nabla_\mu\nabla_\nu \xi -4R_\mu^\lambda\nabla_\lambda \nabla_\nu\xi-{}\\
&{}-4R_\nu^\lambda \nabla_\lambda\nabla_\mu\xi +4R_{\mu\nu}\square\xi +4g_{\mu\nu} R^{\alpha\beta}\nabla_\alpha\nabla_\beta\xi +4R_{\mu\alpha\beta\nu}\nabla^\alpha\nabla^\beta\xi -{}\\
& {}-\frac{1}{2}g_{\mu\nu}(FR+\xi \mathcal{G}-f(R)-\Phi (\mathcal{G}))+(1-F)(R_{\mu\nu} -\frac{1}{2}g_{\mu\nu} R)\,.
		\end{split}\end{equation}
The stress-energy tensor is
\begin{equation}
	T_{\mu \nu}=-\frac{2}{\sqrt{-g}} \frac{\delta S_{M}}{\delta g^{\mu \nu}}\,.
	\end{equation}
Note that here and henceforth we use the notations \mbox{$F(R)=\partial f(R)/ \partial R$}, \mbox{$\xi(\mathcal{G})=\partial \Phi(\mathcal{G})/\partial \mathcal{G}$}.\par 
As the attempt to consider the general $f(R)$ even without the Gauss-Bonnet term \cite{AGPT} appears to be not fully successful and was criticized in \cite{CTD}, we consider a particular simplest and mostly natural family of power-law functions in the Lagrangian.
Namely, we study the case of  $f(R)=R+\alpha R^n$ and $\Phi(\mathcal{G})=\beta \mathcal{G}^m$ because it is important to reproduce GR results in a low-curvature limit.\par 
In the high-curvature limit, the Einstein contribution can be neglected, so we can choose $f(R)=R^n$. 
The Gauss-Bonnet term is a total derivative in $(3+1)$ space-time, therefore, a linear function of $\mathcal{G}$ in the action will not contribute to the field equations.
We consider the Friedman-Lemaitre-Robertson-Walker metric for a flat isotropic universe:
\begin{equation}
ds^2=-dt^2+a^2(t)dx^idx_i\,;
	\end{equation}
using the Hubble parameter \(H={\dot{a}}/{a}\), we  rewrite the field equations in the following form:
\begin{subequations}\label{eq:ab}
	\begin{align}
		3H^2 &= \frac{1}{F}[\frac{1}{2}(FR+\xi \mathcal{G}-f-\Phi)-3H\dot{F}-12H^3\dot{\xi}+8\pi G_N \rho]\,,\label{eq:a}\\ 
		\ddot{F} &= 4\dot{\xi}H^3-4\ddot{\xi}H^2+\dot{F}H-8\dot{\xi}H\dot{H}-8\pi G_N(\rho +p)-2F\dot{H}\,. \label{eq:b}
			\end{align}
				\end{subequations}
Moreover, we will need the continuity equation, the cosmological equation of state and expressions for the Ricci scalar and the Gauss-Bonnet invariant in terms of the Hubble parameter:
\begin{align}
	0 & =\dot{\rho}+3H(\rho +p)\label{eq:cont} \\
	R & =6(2H^2+\dot{H})\label{eq:R} \\
	\mathcal{G} & =24H^2(H^2+\dot{H})\label{eq:G}\\
	p & =\omega \rho \label{eq:state}
		\end{align}
\section{The existence of de Sitter solutions}
The existence of de Sitter solutions, corresponding to an exponentially increasing scale factor (\(H=H_0=const\)), is an important opportunity in modified gravity theories.

Taking the trace of Eq.(\ref{eq:2}), and assuming $F=const, \xi=const , T=0$ we find:
\begin{equation}
	F(R)R+2\xi(\mathcal{G})\mathcal{G}-2f(R)-2\Phi(\mathcal{G})=0\,.
		\end{equation}
Consider the cases, mentioned in the previous section:		
\paragraph{1}\(f(R)=R^n, \Phi(\mathcal{G})=\beta \mathcal{G}^m\):
	\begin{equation}
	 \label{eq:4}
			(n-2)R^n+2\beta(m-1)\mathcal{G}^m=0\,.
				\end{equation}
Using the relations \(R=R_0=12H_0^2,\mathcal{G}=\mathcal{G}_0={R_0^2}/{6}\), we get:
\begin{equation}\label{eq:dSRnGm}
(n-2)R_0^n+2\beta (m-1)\left( \frac{R_0^2}{6} \right)^m=0\,.
	\end{equation}
We can see that de Sitter is absent for the following combinations of $m$ and $n$:
\begin{enumerate}
	\item $n=2$ for any $m$
		\item $m=1$ for any $n$
			\item $n\neq2$ , $m=n/2$
		\end{enumerate}
Thus de Sitter solutions exist for all other combinations of $m$ and $n$.
\paragraph{2}\(f(R)=R+\alpha R^n, \Phi(\mathcal{G})=\beta \mathcal{G}^m\):
\begin{equation}
	\label{eq:5}
	2\beta (m-1)\mathcal{G}^m+\alpha(n-2)R^n-R=0\,
		\end{equation}
and
\begin{equation}\label{eq:dSRRnGm}
	2\beta (m-1)\left(\frac{R_0^2}{6}\right)^m+\alpha(n-2)R_0^n-R_0=0\,.
		\end{equation}
We can see that de Sitter is absent for the following combinations of $m$ and $n$:
\begin{enumerate}
	\item $n=2$, $m=1$
		\item $m=1/2$ and $n=1,2$
		\end{enumerate}
Thus, de Sitter solutions exist for all other combinations of $m$ and $n$.\par
We can see that adding $\Phi(\mathcal{G})$ function into the modified gravity action allows us to obtain new de Sitter solutions  which do not 
exist in	\(f(R)\) and pure Gauss-Bonnet gravity.\par 					
\section{Cosmological dynamics}
Since the field equations in Modified Gravity theories are non-linear fourth order differential equations, which are  very difficult to solve, we use the Dynamical system approach. It provides a powerful and relativity simple scheme for obtaining asymptotic solutions and investigating their stability.\par  
The main goal of this approach is to get some autonomous system of first order differential equations, as a consequence of the cosmological equations of motion. In this approach, the dynamics of the Universe corresponds to motion along a phase curve, and stationary points represent some asymptotic regimes of the Universe evolution.\par 
Dividing Eq. (\ref{eq:a}) by \(3H^2\), we find:
	\begin{equation}\label{eq:oo}
1= \frac{R}{6H^2}+\frac{\xi \mathcal{G}}{6FH^2}-\frac{f}{6FH^2}-\frac{\Phi}{6FH^2}-\frac{\dot{F}}{FH}-\frac{4H\dot{\xi}}		{F}+\frac{8\pi G_N \rho}{3FH^2}\,.
		\end{equation}
Now we  introduce the following normalized expansion variables:
\begin{subequations}\label{eq:sys1}
\begin{align}
	x & =\frac{\dot{F}}{FH}\,,\\
	y & =\frac{R}{6H^2}\,,\\
	z & =\frac{f}{6FH^2}\,,\\
	\Omega & =\frac{8\pi G_N \rho}{3FH^2}\,,\\
	\phi & =\frac{\xi \mathcal{G}}{6FH^2}\,,\\
	\kappa & =\frac{\Phi}{6FH^2}\,,\\
	\psi & =\frac{4H\dot{\xi}}{F}\,.
		\end{align}
				\end{subequations}
We take derivative of these variables with respect to the dimensionless time \(N=|\ln a(t)|\): 				
\begin{subequations}\label{sys:dynamicx}
	\begin{align}
		\frac{dx}{dN} & =\frac{\ddot{F}}{FH^2}-x^2-x\frac{\dot{H}}{H^2}\,,\\
		\frac{dy}{dN} & =\frac{\dot{R}}{6H^3}-2y\frac{\dot{H}}{H^2}\,,\\
		\frac{dz}{dN} & =\frac{\dot{R}}{6H^3}-zx-2z\frac{\dot{H}}{H^2}\,,\\
		\frac{d\Omega}{dN} & =\frac{8\pi G_N\dot{\rho}}{3FH^3}-\Omega x-2\Omega \frac{\dot{H}}{H^2}\,,\\
		\frac{d\psi}{dN} & =\frac{4\ddot{\xi}}{F}+\psi \frac{\dot{H}}{H^2}-\psi x\,,\\
		\frac{d\phi}{dN} & =\psi \frac{\mathcal{G}}{24H^4}+\phi \frac{\dot{\mathcal{G}}}{\mathcal{G}H}-\phi x-2\phi \frac{\dot{H}}{H^2}\,,\\
		\frac{d\kappa}{dN} & = \phi \frac{\dot{\mathcal{G}}}{\mathcal{G}H}-\kappa x-2\kappa \frac{\dot{H}}{H^2}\,.
			\end{align}
				\end{subequations}
Using the new variables,  Eq.(\ref{eq:oo}) can be rewritten as
\begin{align}
1 & = y+\phi- z-\kappa -x-\psi+\Omega \,. \label{constraint}
	\end{align}
We have obtained a set of seven equations, but our theory is of the fourth order. As we will see further, our dimensionless variables are not independent, consequently, this system is overdetermined, and our next purpose is to reduce its dimensionality.\par 
To complete transformation, we need to express all right-hand side terms in Eqs. (\ref{sys:dynamicx}) through the variables defined in Eqs.(17).\\
Using Eqs. (\ref{eq:R}), (\ref{eq:cont}), (\ref{eq:G}), and (\ref{eq:state}), we find:
\begin{align}
\frac{\dot{H}}{H^2} & =y-2\,, \\
\frac{\dot{R}}{6H^3} & =\frac{xy}{b}\,,\\
\frac{8\pi G_N\dot{\rho}}{3FH^3} & =-3\Omega (1+\omega)\,,\\
\frac{\mathcal{G}}{24H^4} & =y-1\,,\\
\frac{\dot{\mathcal{G}}}{\mathcal{G}H} & =\frac{(2(y-2)^2+\frac{xy}{b})}{y-1}\,,
\intertext{where} b & =\frac{f_{,RR} R}{F}, \quad f_{,RR}=\frac{\partial ^2f(R)}{\partial R^2}\,.
\end{align}
If the functions $f(R)$ and $\Phi(\mathcal{G})$ are chosen, we can close the system due to additional relations between our variables.

\textbf{The case of \(\Phi (\mathcal{G})=\beta \mathcal{G}^m\). }
Using Eqs. (\ref{eq:R}) and (\ref{eq:G}), we get:
\begin{align}
	\phi & =m \kappa\,, \\
	\psi & =\frac{\phi (m-1)(2(y-2)^2+\frac{xy}{b})}{(y-1)^2}\,.\label{eq:psiphi} 
		\end{align}
Thus, using these relations with the constraint (\ref{constraint}), we exclude $\kappa$,$\psi$ and $x$ from our autonomous system.
As a result, we obtain a set of four equations:
\begin{subequations}\label{sys:final}\begin{align}
	\frac{dy}{dN} & =\frac{xy}{b}-2y(y-2)\,,\\
	\frac{dz}{dN} & =\frac{xy}{b}-zx-2z(y-2)\,,\\
	\frac{d\Omega}{dN} & =-3\Omega (1+\omega)-\Omega x-2\Omega (y-2)\,,\\
	\frac{d\phi}{dN} & = \phi \frac{ m(2(y-2)^2+\frac{xy}{b})}{(y-1)}-\phi x-2\phi (y-2)\,,
	\end{align}\end{subequations}
where 
\begin{align}
	x & =\frac{\Omega -1+y-z+\phi (m-1)(\frac{1}{m}-\frac{2(y-2)^2}{(y-1)^2})}{1+\frac{y\phi (m-1)}{b(y-1)^2}}\,,\\
	b & =\frac{f_{,RR} R}{F}\,.
		\end{align}
We would get  additional relations between the variables for every particular function  \(f(R)\).\\
\section{The high-curvature case of \(f(R)=R^n\)} 
First of all, we can easily find in this special case
\begin{align}
y & =nz\,,\\
b & =n-1=const\,.
\end{align}
We can substitute these relations into (\ref{sys:final}) and exclude $z$ from our system:
\begin{align}\begin{split}
	\frac{dy}{dN} & =\frac{xy}{n-1}-2y(y-2)\,,\\
	\frac{d\Omega}{dN} & =-3\Omega (1+\omega)-\Omega x-2\Omega (y-2)\,,\\
	\frac{d\phi}{dN} & =\phi\frac{ m(2(y-2)^2+\frac{xy}{n-1})}{y-1}  -\phi x-2\phi(y-2)\,,\\
	x & =\frac{\Omega -1+y\frac{n-1}{n}+\phi (m-1)(\frac{1}{m}-\frac{2(y-2)^2}{(y-1)^2})}{1+\frac{y\phi (m-1)}{(n-1)(y-1)^2}}\,.\\
		\end{split}
			\end{align}
Note that the factor $n-1$ is a denominator in our system. It means that the selected set of variables is useless for Gauss-Bonnet gravity with the Einstein term, where $n=1$.

Solving the system with vanishing left-hand sides, we find the following fixed points $(x,y,\Omega ,\phi)$:
\begin{center}
	\((-1,0,0,0)\)\,,\\
	\((1-3\omega ,0,2-3\omega ,0)\)\,,\\
	\((-3\frac{(n-1)(1+\omega)}{n},\frac{4n-3(1+\omega) }{2n},-\frac{8n^2-13n+3+\omega (6n-3)(n-1)}{2n^2},0)\)\,,\\
	\((\frac{2(n-2)}{2n-1},\frac{(4n-5)n}{(n-1)(2n-1)},0,0)\)\,,\\
	\((0,2,0,-\frac{(n-2)m}{n(m-1)})\)\,,\\
	\((-8m+4, 0, 0,\frac{m(8m-5)}{(m-1)(8m-1)})\)\,.
\end{center}
The first four points are precisely the same as those obtained by Carloni, Dunsby, Capozziello and Troisi for \(R^n\)-gravity \cite{CDCT}.\footnote[1]{ Note that they have used a bit different set of variables and signature from our paper.} The Gauss-Bonnet term does not contribute here. 
Except the $R^n$-gravity solutions, here are two new stationary points:\\
\begin{enumerate}
	\item \((0,2,0,-\frac{(n-2)m}{n(m-1)})\).
It is the de Sitter point.
	\item \((-8m+4, 0, 0,\frac{m(8m-5)}{(m-1)(8m-1)})\).
The scale factor is \(a(t)=a_0(t-t_0)^\frac{1}{2}\), so this point represents the well-known scale factor evolution dynamics of $f(R)$-gravity,
though
the set of coordinates for this point is different from the first and second points, which also correspond to $a\sim t^{\frac{1}{2}}$.
		\end{enumerate}
		
It was shown in \cite{STT} (where combined effects of $f(R)$ and $R \Box R$ corrections were studied)  
that using the dynamical system approach we can obtain some new non-trivial solutions mainly where the correction terms in the 
action are equally important for a power-law evolution of the scale factor. Also there exist some additional relations between variables in this case. As we see further, the similar picture exists in our system. 

\textbf{The case \(m={n}/{2}\) in $R^n+\beta \mathcal{G}^m$ models.}
There is an additional relation between $\phi$ and $y$ in this special case: 
\begin{equation}
\phi = \beta \frac{2^\frac{n}{2}}{3^\frac{n}{2}}\frac{(y-1)^\frac{n}{2}}{2y^{n-1}}\,.
\end{equation}
Consequently, one can reduce the system to two equations:
\begin{align}\begin{split}\label{sys:RnGn/2}
\frac{dy}{dN}  = &\Bigg[ \Omega -1+y-\frac{y}{n}+\beta \frac{2^\frac{n}{2}}{3^\frac{n}{2}}\frac{(y-1)^\frac{n}{2}}{2y^{n-1}} (\frac{n}{2}-1)(\frac{2}{n}-\frac{2(y-2)^2}{(y-1)^2})\Bigg] {}\\
&\times{\Bigg[\frac{n-1}{y}+\frac{\beta(\frac{n}{2}-1)2^{\frac{n}{2}-1}(y-1)^{\frac{n}{2}-2}}{3^{\frac{n}{2}}y^{n-1}}\Bigg]^{-1}}-2y(y-2)\,,\\
\frac{d\Omega}{dN}  =& -\Omega(3(1+\omega)+2(y-2))-{}\\
& \Omega \Bigg[\Omega -1+y-\frac{y}{n}+\beta \frac{2^\frac{n}{2}}{3^\frac{n}{2}}\frac{(y-1)^\frac{n}{2}}{2y^{n-1}} (\frac{n}{2}-1)(\frac{2}{n}-\frac{2(y-2)^2}{(y-1)^2})\Bigg] {}\\
&\times {\Bigg[1+\frac{\beta(\frac{n}{2}-1)2^{\frac{n}{2}-1}(y-1)^{\frac{n}{2}-2}}{3^{\frac{n}{2}}y^{n-2}(n-1)}\Bigg]^{-1}}\,.
\end{split}
\end{align}
\par 
It is useful to describe our future strategy. We have started from the model \mbox{$R+\alpha R^n+\beta \mathcal{G}^m$} and consider its high-curvature limit, where $R$ can be neglected.  We have two coupling constants, $\alpha$ and $\beta$, however, when neglecting the Einstein term in the action, effectively our result depends only on their ratio $\beta$/$\alpha$. Thus we can assume $\alpha =1$.  Next, we  select the model by fixing $n$ and find stationary points for some values of $\beta$.  Having this information we will discuss the situation in the general case. \par 
We mostly investigate the case of positive integers $n$ and $m$, so $n$ is chosen to be even. If $n=2$, $m=1$ so the Gauss-Bonnet term does not contribute. Thus, we study the case of $n=4$ in detail.

\textbf{The case of $n=4$,$\beta=1$.}
Stationary points in this model, which have been found numerically are, by the scheme ($x$, $y$, $\Omega$, $\phi$):
\begin{center}
$\mathcal{A}1$:($-9.358$,\quad $0.440$,  \quad  $0$,\quad $0.816$)\,,\\
\; $\mathcal{B}1$:($0.576$,  \; \; $2.096$, \quad \, $0$,\quad  $0.028$)\,,\\
$\tiny{\mathcal{C}1:\left(-\frac{9}{4}-\frac{9}{4}\omega, \frac{13}{8}-\frac{3}{8}\omega ,-\frac{1}{288}\frac{15309\omega^4 -179820\omega^3 +628974\omega^2 -168876\omega- 1599187}{(3\omega -13)^3}, \frac{2(0.625-0.375\omega)^2}{9(1.625-0.375\omega)^3} \right)}$ \,.\\
\end{center}
The corresponding eigenvalues are (for the last point eigenvalues depends on equation-of-state parameter $\omega$, so the values are presented for several particular $\omega$):
\begin{align*}
\mathcal{A}1&:(7.918,\quad \; \, \; 9.478-3\omega)\,,\\
\mathcal{B}1&:(-3.672, \; -3.768-3\omega)\,,\\
\mathcal{C}1&:\left\{ \begin{aligned}
\omega &=-1, \quad & 0.722,\quad & -3.722 \\    
\omega &=-0.5,\quad & 1.965,\quad & -3.653 \\ 
\omega &=0,\quad & 3.081,\quad & -3.456\\
\omega &=0.5,\quad & 4.115,\quad & -3.177\\
\omega &=1,\quad & 5.083,\quad & -2.833\\    
\end{aligned} \right\}
\,.
\end{align*}
These points are summarized in the Table 1.
\begin{table}[h!]
\caption{\label{tab:R4G2} Stationary points for $R^4+\mathcal{G}^2$}
	\begin{center}
		\begin{tabular}{|c|c|c|c|}
			\hline
			Point & Coordinates $(x , y , \Omega , \phi)$ & Stability  & The scale  \\ 
			& 		of stationary point							& type & factor, $a(t)$ \\ \hline 
			$\mathcal{A}1$ & $x =-9.358$  &  &  \\ 
  & $y =0.440$ & Repulsive node & $a_0(t-t_0)^\frac{1}{1.560}$ \\ 
  & $\Omega =0$ &  &  \\ 
  & $\phi =0.816$ &  &  \\ \hline
$\mathcal{B}1$ & $x =0.576$  &  &  \\ 
  & $y =2.096$ & Attractive node & $a_0(t-t_0)^{-\frac{1}{0.096}}$ \\ 
  & $\Omega =0$ &  &  \\ 
  & $\phi =0.028$ &  &  \\ \hline
$\mathcal{C}1$ & $x =-\frac{9}{4}-\frac{9}{4}\omega$  &  &  \\ 
  & $y =\frac{13}{8}-\frac{3}{8}\omega$ & Saddle & $a_0(t-t_0)^{\frac{8}{3(\omega+1)}}$ \\ 
  & \tiny{ $\Omega=-\frac{15309\omega^4 -179820\omega^3 +628974\omega^2 -168876\omega- 1599187}{288(3\omega -13)^3}$} &  &  \\ 
  & $\phi =\frac{2(0.625-0.375\omega)^2}{9(1.625-0.375\omega)^3}$ &  &  \\ \hline
	\end{tabular}
		\end{center}
			\end{table}
			
\textbf{The case of $n=4$, $\beta=-1$.}
Stationary points in this model are, by the scheme $(x, y, \Omega, \phi)$:
\begin{center}
$\mathcal{A}2:$($-15.754$,    $-0.625$,  \quad \, $0$, \quad   $2.396$)\,,\\
\quad \;$\mathcal{B}2:$($0.566$, \quad  \; $2.094$, \quad \! $0$,  \,  $-0.028)$\,,\\
$\tiny {\mathcal{C}2:\left(-\frac{9}{4}-\frac{9}{4}\omega, \frac{13}{8}-\frac{3}{8}\omega ,-\frac{1}{288}\frac{15309\omega^4 -179820\omega^3 +596718\omega^2 -159660\omega- 1524947}{(3\omega -13)^3}, -\frac{2(0.625-0.375\omega)^2}{9(1.625-0.375\omega)^3} \right)}$\,. \\
\end{center}
The stability types for all this points did not change as compared with the case $\beta =1$. 
The situation in all these cases is the same as has been found in $R^n$ gravity -- only one stationary point (which corresponds to a phantom solution) is stable, others are unstable.

\textbf{The general case of $n=4$.}
Now we try to describe what happens for an arbitrary $\beta$.
Substituting $a(t)=a_0(t-t_0)^{\frac{1}{A}}$ into Eq. (\ref{eq:b}), we get for $R+\alpha R^4+\beta \mathcal{G}^{2}$ ($n=4$) an algebraic instead of differential equation:
\begin{equation}\label{eq:1/A}\begin{split}
&-\frac{1}{A^7(t-t_0)^8}(248832 \alpha A-414720 \alpha A^2+214272 \alpha A^3-36288 \alpha A^4+{}\\
& +768\beta+4608\beta A-5376\beta A^2+27648\alpha -{}\\
&-2A^6(t-t_0)^6+8\pi G_N \rho A^7(t-t_0)^8(1+\omega))=0\,.
	\end{split}
		\end{equation}
Neglecting matter and the Einstein term, assuming $\alpha =1$, we find the equation for $A$ as a function of $\beta$:
\begin{equation} 
-144-1296A+2160A^2-1116A^3+189A^4-4\beta -24\beta A+28 \beta A^2=0\,.
	\end{equation}
\begin{figure}[h!]
	\includegraphics[width=150mm]{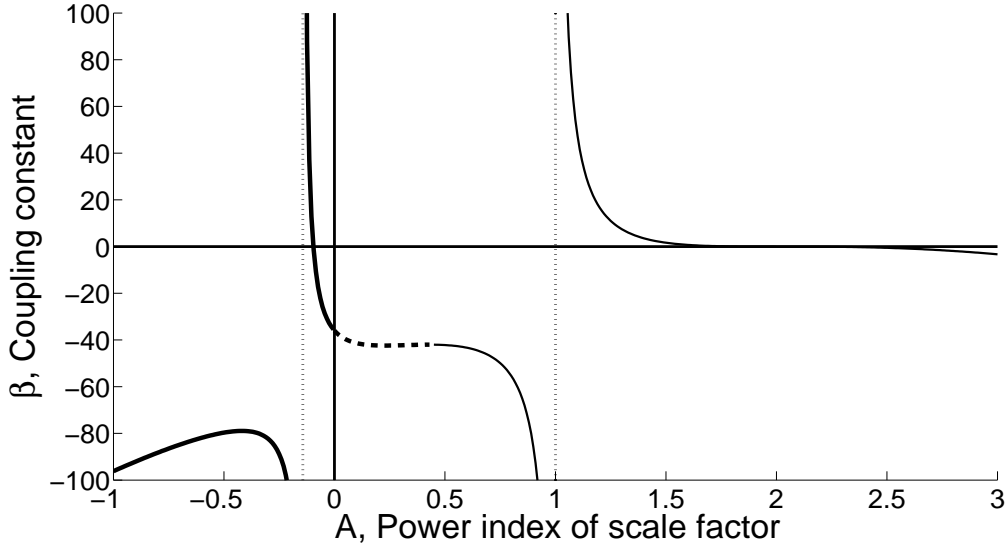}
	\caption{Power index A of a vacuum power-law solutions depending on coupling constant for $\mathcal{L}=R^4+\beta \mathcal{G}^2$. Stable branches are solid bold lines, branches with stability depending on $\omega$ are dashed bold lines.}
		\end{figure}
\par A Stability analysis provided for a set of particular $\beta$ reveals the following picture: 
\begin{itemize}
\item A$<$0. In this area there are only phantom branches, which are always stable. For $\beta =\pm 1$ they are solutions $\mathcal{B}1,2$. Note that in the coupling constant interval {$-36>\beta \gtrsim
-78.944$} we do not find any phantom solution at all.
\item{A$>$1}. Branches of this region are always unstable.
\item{A$\in[0,3/7 \approx 0.428]$}. For this interval $\beta \in$[$-36$,$\approx-42.402$] the branch stability depends on the parameter $\omega$ (e.g., for $\beta=-36$ the eigenvalues are \mbox{[$-3,-3-3\omega$]}). 
Considering also the matter-dominated solution, we get the following picture:
there is some critical value 
\begin{equation} 
\omega^*=\frac{8A-3}{3}\,.
\end{equation}
In the case of $\omega<\omega^*$, the matter-dominated solution is stable while the vacuum power-law solution is unstable. In the opposite case $\omega>\omega^*$, the solution with $a\sim t^{{1}/{A}}$ is stable while $a\sim t^{{8}/{[3(1+\omega)]}}$ is unstable, i.e.  solution with a larger power index is always stable. 
Thus, there is always one stable non-phantom solution for any $\omega$ in the range of the coupling constants $-42.018 \lesssim
\beta <-36$. In the range $-42.402\lesssim \beta \lesssim -42.018$ (between the local maximum and minimum of function $\beta(A)$, see Fig. 2) we can find two stable vacuum solutions for some values of $\omega$.
\item{A$>3/7\sim 0.428$}. Here the vacuum power-law solution becomes unstable for any $\omega$, while the matter-dominated solution $a\sim t^{{8}/{[3(1+\omega)]}}$ is still stable for some $\omega$ values close to $-1$. Furthermore, it becomes also unstable for $\beta \lesssim -54, A \gtrsim 0.791 $.

However, it does not mean that there are no stable solutions for this range of coupling constants. In this section we considered models without the Einstein term, i.e., high-curvature regimes. Actually, earlier we noted that we should add  $R$ contribution for obtaining a late-time standard cosmology. This contribution  gives us a de Sitter solution, which is stable in this range of coupling constants (see below).
\end{itemize}
\begin{figure}[h!]
	\begin{center}
	\includegraphics[width=150mm]{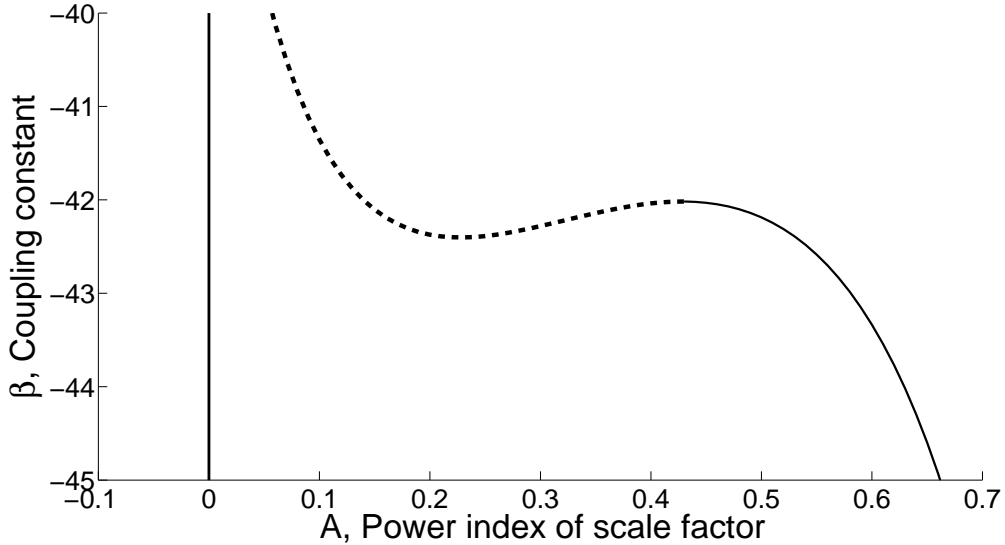} 
	\caption{One of the branches plotted in Fig.1 is shown here in more detail . The notation are the same as in Fig.1.}
		\end{center}
		\end{figure}\par
\textbf{The  case of general  n.}
Note, that Eq.(\ref{eq:1/A}) can be simplified only when the power-law ansatz terms originating from $f$ and $\Phi$ contribution are equally important. In this section we considered the case $n=4$ for $R^n+\beta \mathcal{G}^{\frac{n}{2}}$ and discussed it in detail. Our studies show that the picture obtained does not change qualitatively  for any  $n>2$. In all these models we can consider $\beta$ as a function of the power index in the time dependence of the scale factor $A$,
\begin{equation}
\beta(A)=-2\frac{(A-1)(-24(A-1))^{-\frac{n}{2}}(-6(A-2))^n(n-3nA-2+A+2n^2A)}{(n-5nA-2+2A+2An^2)(A-2)}
	\end{equation}
	 and admit some common features:
\begin{enumerate} 
	 \item There is a stable non-phantom solution in the coupling constants range \mbox{$\beta_{min} \leq \beta \leq\beta(0)$}, where $\beta_{min}$ is the value of the function $\beta (A)$ at the local minimum point 
\[
A=\frac{1}{2}\frac{-n+\sqrt{17n^2-24n+8}}{n^2-3n+1}\,,
\]
$\beta(A=0)=-(6)^{\frac{n}{2}}$. Moreover, there is some critical value $\omega^*$, determining stability in this range:
	\begin{equation}
	\omega^* =\frac{2nA-3}{3}
	\end{equation}
	Solution with a larger power index (it is a matter dominated point or a vacuum power-law solution, depending on $\omega$) is stable.
	 \item After $A$ reaches the value $A={3}/{(2n-1)}$, this branch becomes unstable. 
	 \item There also exists a range of the coupling constants 
	 \[
	 \beta(-\frac{1}{2}\frac{n+\sqrt{17n^2-24n+8}}{n^2-3n+1})<\beta <\beta(0), 
	 \]
	 where we have no phantom solutions at all. 
	 \end{enumerate} 
	 The only qualitative difference for the case of odd and non-integer $n$ is the absence of $\mathcal{A}$-type solution in the range $A>1$.\par
We should conclude that for $\beta =-(6)^{\frac{n}{2}}$ there is a stable solution with $A=0$. It means that this solution cannot be of power-law type.
Using Eq. (\ref{eq:dSRnGm}) we easily find that it cannot be a de Sitter point. 
Since in Quantum field theory coupling constants are usually running, we do not consider in detail this solution existing only for a particular
value of the coupling constant.

It is interesting that in $R^n$-gravity with $n>2$ we always have a stable phantom solution. By using the power-law ansatz it is simple
to obtain that in pure Gauss-Bonnet gravity with $\mathcal{F} (R,\mathcal{G})=\mathcal{G}^m$ there is a phantom solution when $m>1/4$. So, a
combined effect of both $R^n$ and $\mathcal{G}^m$ correction terms may lead to solutions that are qualitatively different from those
in pure $R^n$ or $\mathcal{G}^m$ theories.

\newpage 
\section{Dynamics in \(R+\beta \mathcal{G}^m\) theories}
Now we consider cosmological dynamics in a theory where the only correction to Einstein gravity is a function of the Gauss-Bonnet term.
It was shown in Sec.5 that the set of dimensionless variables (\ref{eq:sys1}) does not allow us to study this case, so we choose another set of variables. Substituting $f(R)=R$ into (\ref{eq:ab}), we find the field equations: 
\begin{align}\label{eq:pureGaussbonnet}
3H^2 &= \frac{1}{2}(\xi \mathcal{G}-\Phi)-12H^3\dot{\xi}+8\pi G_N \rho\,,\\
	0 &= 4\dot{\xi}H^3-4\ddot{\xi}H^2-8\dot{\xi}H\dot{H}-8\pi G_N(\rho +p)-2\dot{H}\,.
		\end{align}
We rewrite these equations in a dimensionless form:
\begin{align}
	1 &= \frac{\xi \mathcal{G}}{6H^2}-\frac{\Phi}{6H^2}-4H\dot{\xi}+\frac{8\pi G_N \rho}{3H^2}\,.\\
	0 &= 4\dot{\xi}H-4\ddot{\xi}-8\dot{\xi}H\frac{\dot{H}}{H^2}-\frac{8\pi G_N(\rho +p)}{H^2}-2\frac{\dot{H}}{H^2}\,.
		\end{align}
Substituting \(\Phi =\beta \mathcal{G}^m\), \(\xi \mathcal{G}=m\beta \mathcal{G}^m\), we get:
\begin{align}
	1 &= \beta \frac{(m-1) \mathcal{G}^m}{6H^2}-4H\dot{\xi}+\frac{8\pi G_N \rho}{3H^2}\,.
		\end{align}
The new dimensionless variables are
\begin{subequations}
	\begin{align}
	\phi & =\beta \frac{(m-1) \mathcal{G}^m}{6H^2}\,,\\
	\psi & =4H\dot{\xi}\,,\\
	\Omega & =\frac{8\pi G_N \rho}{3H^2}\,,\\
	\sigma & =\frac{\dot{H}}{H^2}\,.
		\end{align}
				\end{subequations}
Taking derivative of the introduced variables with respect to $N=|\ln a(t)|$, we get the following set of equations:
\begin{subequations}
	\begin{align}
		\frac{d\Omega}{dN} & =-3\Omega (1+\omega)-2\Omega \sigma\,,\\
			\frac{d\psi}{dN} & =\psi \sigma +4\ddot{\xi}\,,\\
			\frac{d\phi}{dN} & =m\phi \frac{\dot{\mathcal{G}}}{\mathcal{G}H}-2\phi \sigma\,,\\
			\frac{d\sigma}{dN} & =\frac{\ddot{H}}{H^3}-2\sigma ^2\,.
				\end{align}
					\end{subequations}
Taking into account the relations
\begin{align}
	0 & = \psi -4\ddot{\xi}-2\psi \sigma -3\Omega (1+\omega)-2\sigma\,,\\
	1 & =\phi -\psi +\Omega\,,\\
	\frac{\dot{\mathcal{G}}}{\mathcal{G}H} & =\frac{\psi (1+\sigma)}{m\phi}\,,\\
	\frac{\ddot{H}}{H^3} & =\frac{\psi (1+\sigma)^2}{m\phi}-4\sigma -2\sigma ^2\,,
		\end{align}
we find the resulting autonomous system:
\begin{subequations}
	\begin{align}
	\frac{d\Omega}{dN} & =-3\Omega (1+\omega)-2\Omega \sigma\,,\\
	\frac{d\phi}{dN} & =(\phi -1 +\Omega)(1+\sigma)-2\phi \sigma\,,\\
	\frac{d\sigma}{dN} & =(\phi -1 +\Omega) \frac{(1+\sigma)^2}{m\phi}-4\sigma -4\sigma ^2\,.
		\end{align}	
			\end{subequations}
Equating the left-hand side of our system to zero, we find the stationary point $(\phi=1, \psi=0, \Omega=0, \sigma=0)$, which corresponds to a de Sitter solution.

As in the previous case, nontrivial results have been obtained for a particular value of the power index $m$.

\textbf{The case $m=\frac{1}{2}$.}
There is an additional relation in the case of $m=\frac{1}{2}$:
\begin{equation}
	\phi = -\beta \sqrt{\frac{\sigma +1}{6}}\,.
		\end{equation}
We can substitute this relation into our system and find
\begin{subequations}\begin{align}
	\frac{d\Omega}{dN} & =-3\Omega (1+\omega)-2\Omega \sigma\,, \\
	\frac{d\sigma}{dN} & = 2 \frac{\sqrt{6}(1-\Omega)+\beta \sqrt{\sigma +1}}{\beta}(1+\sigma)^{\frac{3}{2}}-4\sigma -4\sigma ^2\,.
		\end{align}\end{subequations}
Using the technique applied in the previous section, we find stationary points for particular values of $\beta$.
For $\beta =1$ they are, by the scheme ($\phi, \psi, \Omega,\sigma$):
\begin{center}
$\mathcal{A}3:(0, -1,\quad  0, -1)$\,,\\
$\mathcal{B}3:(-1.263, -2.263,\quad  0,  \quad  8.582$)\,, \\
$\mathcal{C}3:(-\frac{1}{2}\sqrt{-\frac{1}{3}-\omega},-\frac{\sqrt{-9-3\omega}(1+\omega)}{3\omega +1},\frac{1}{6}\frac{6(3\omega +1)-(5+3\omega)\sqrt{-3-9\omega}}{3\omega +1}, -\frac{3}{2}-\frac{3}{2}\omega$)\,.
\end{center}
The corresponding eigenvalues are:
\begin{align*}
	\mathcal{A}3&:(4,\quad  -3\omega -1)\,,\\
	\mathcal{B}3&:(-11.582,-3\omega -20.165)\,,\\
	\mathcal{C}3&:\left\{ \begin{aligned}
	\omega &=-1, \quad & 2.505,\quad & -5.505 \\    
	\omega &=-0.8,\quad & 1.996,\quad & -4.696 \\ 
	\omega &=-0.6,\quad &  1.345,\quad & -3.745\\
	\omega &=-0.4,\quad & 0.423,\quad & -2.523\\  
		\end{aligned} \right\}\,.\\
			\end{align*}
Now turn to the case of  $\beta =-1$. We get the fixed points by the scheme ($\phi, \psi, \Omega,\sigma$):
\begin{center}
	$\mathcal{A}3:(0, -1,\quad  0, -1)\,,$\\
	$\mathcal{D}3:(0.263, -0.736,\quad  0,\quad  -0.582$)\,,\\
	$\mathcal{E}3:(\frac{\sqrt{-3-9\omega}}{6},\frac{\sqrt{-9-3\omega}(1+\omega)}{3\omega +1},\quad \frac{1}{6}\frac{6(3\omega 		+1)+(5+3\omega)\sqrt{-3-9\omega}}{3\omega +1}, -\frac{3}{2}-\frac{3}{2}\omega$)\,,
	\end{center}
with the eigenvalues
\begin{align*}
\mathcal{A}3:&(4,\quad  -3\omega -1)\,,\\
\mathcal{D}3:&(-2.417,\quad -3\omega -1.834)\,,\\
\mathcal{E}3:&\left\{ \begin{aligned}
\omega &=-1, \quad &-1.5-1.883i,\quad & -1.5+1.883i \\    
\omega &=-0.8,\quad & -1.350+0.525i,\quad & -1.350-0.525i\\ 
\omega &=-0.6,\quad & 0.033,\quad & -2.433\\
\omega &=-0.4,\quad & 0.196,\quad & -2.296\\    
		\end{aligned} \right\}\,.\\
			\end{align*}
			We summarize our results in Table 2.
\begin{table}[h!]
\caption{\label{tab:RGm} Fixed points for $R+\beta \sqrt{\mathcal{G}}$}
\begin{center}
\begin{tabular}{|c|c|c|c|}
\hline
Point & Coordinates $(\phi , \psi , \Omega , \sigma)$ & Stability      & The scale  \\ 
      &  of stationary points                         &  type                & factor, $a(t)$\\ \hline
      \multicolumn{4}{|c|}{$R+\sqrt{\mathcal{G}},\quad \beta=1$} \\ \hline
$\mathcal{A}3$ & $\phi =0$  &  &  \\ 
  & $\psi =-1$ & Repulsive node & $a_0(t-t_0)$ \\ 
  & $\Omega =0$ &  &  \\ 
  & $\sigma =-1$ & &  \\ \hline
$\mathcal{B}3$ & $\phi =-1.263$ &   & \\
  & $\psi= -2.263$ & Attractive node & $a_0(t-t_0)^{-\frac{1}{8.582}}$\\
  & $\Omega =0$ &  & \\
  & $\sigma = 8.582$ &  & \\ \hline 
$\mathcal{C}3$  & $\phi =-\frac{\sqrt{-3-9\omega}}{6}$ &  & \\
  & $\psi =-\frac{\sqrt{-9-3\omega}(1+\omega)}{3\omega +1}$ & Saddle & $a_0(t-t_0)^{\frac{2}{3(1+\omega)}}$\\
  & $\Omega =  \frac{1}{6}\frac{6(3\omega +1)-(5+3\omega)\sqrt{-3-9\omega}}{3\omega +1}$ & & \\
  & $\sigma = -\frac{3}{2}-\frac{3}{2}\omega$ &  & \\ \hline
        \multicolumn{4}{|c|}{$R-\sqrt{\mathcal{G}},\quad \beta=-1$} \\ \hline
$\mathcal{A}3$ & $\phi =0$  &  &  \\ 
  & $\psi =-1$ & Repulsive node & $a_0(t-t_0)$ \\ 
  & $\Omega =0$ &  &  \\ 
  & $\sigma =-1$ &  &  \\ \hline
$\mathcal{D}3$ & $\phi =0.263$  & $\omega<-0.611$  &  \\ 
  & $\psi =-0.736$ & Saddle & $a_0(t-t_0)^\frac{1}{0.582}$ \\ 
  & $\Omega =0$ & $-1/3>\omega>-0.611$ &  \\ 
  & $\sigma =-0.582$ & Attractive node &  \\ \hline
$\mathcal{E}3$  & $\phi =\frac{\sqrt{-3-9\omega}}{6}$ & & \\
  & $\psi =\frac{\sqrt{-9-3\omega}(1+\omega)}{3\omega +1}$ & $ \omega <-0.777$, Attractive focus  & $a_0(t-t_0)^{\frac{2}{3(1+\omega)}}$\\
  & $\Omega = \frac{1}{6}\frac{6(3\omega +1)+(5+3\omega)\sqrt{-3-9\omega}}{3\omega +1}$ &  $-0.777< \omega <-0.611$, Attractive node  & \\
  & $\sigma = -\frac{3}{2}-\frac{3}{2}\omega$ & $-0.611< \omega<-1/3$, Saddle  & \\
  &   &  & \\ \hline
		\end{tabular}
			\end{center}
				\end{table}
\newpage
Using the power-law ansatz $a\sim (t-t_0)^{\frac{1}{A}}$ with the field equations (\ref{eq:pureGaussbonnet}) we can investigate the general case:
\begin{equation}
-\frac{1}{2A^2\sqrt{1-A}(t-t_0)^2}(6\sqrt{1-A}+\beta \sqrt{6}(A+1))=0\,.
\end{equation}
In this way one can express the coupling constant $\beta$ as a function of $A$ \mbox{(see Fig 3.)}
\begin{figure}[h!]
\begin{center}
	\includegraphics[width=150mm]{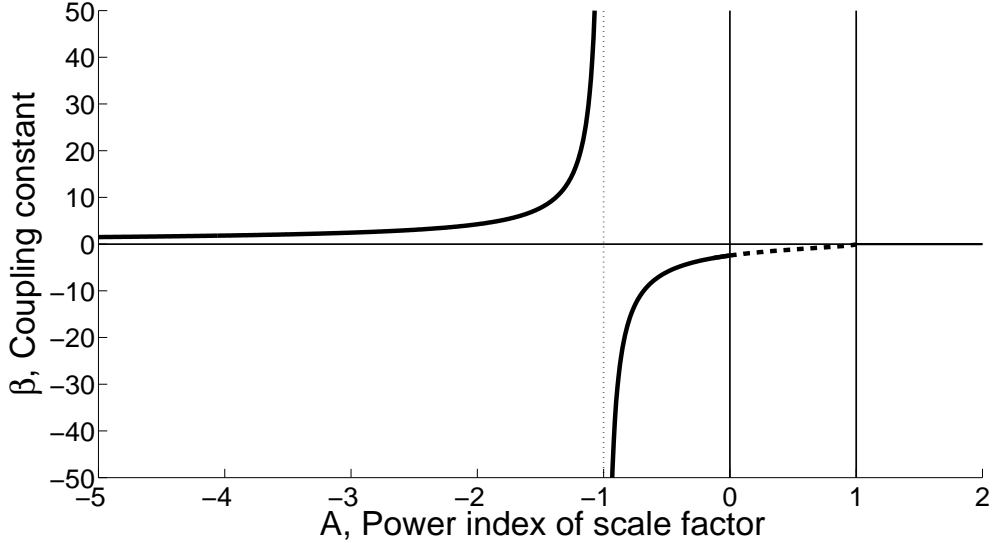} 
	\caption{Power index A of a vacuum power-law solutions depending on coupling constant for $R+\beta \sqrt{\mathcal{G}}$. The notations are the same as in Fig.1}
	\end{center}
		\end{figure}
Incidentally, when $\beta >0$ or $\beta<-\sqrt{6}$, only phantom solutions are stable. When \mbox{$-\sqrt{6}<\beta<0$,} there is only one stable non-phantom solution. It is either a vacuum power-law solution or a matter dominated point. \par 
Similarly to the situation considered in the previous section, there is a critical value   
\begin{equation}
\omega^*=\frac{2A-3}{3}
\end{equation}
that determines the  power-law solutions' stability in this coupling constant range. 
It means that in this theory, like that considered in the previous section, there is a coupling constant range in which we have no vacuum
phantom solutions.

It should be noted that to consider a matter source with $w>-1/3$ we have to change the sign of the Gauss-Bonnet term, i.e., to consider
a theory with $R+\beta \sqrt{-\mathcal{G}}$. In such a theory we have a stable matter dominated point corresponding to $a \sim t^{2/[3(1+\omega)]}$.

\section{Stability of de Sitter solution}
Let us discuss the stability of de Sitter solution. It is the case in which one can easily obtain a stability condition. Consider linear perturbations about the de Sitter point:
\begin{align}
H &=H_0+\delta H\,, &\\
R &=R_0+6(4H_0\delta H +\delta \dot{H}),
&R_0 =12H_{0}^2\,,\\
\mathcal{G} &=\mathcal{G}_0+24(4H_{0}^3\delta H+H_0^2\delta \dot{H}),
&\mathcal{G}_0 =24H_{0}^4\,, \\
\delta \dot{R} &=6(4H_0\delta \dot{H}+\delta \ddot{H})\,,&\\
\delta \dot{\mathcal{G}} &=24(4H_{0}^3 \delta{\dot{H}}+H_{0}^2\delta {\ddot{H}})\,.&
\end{align}
Substituting these relations into Eq. (\ref{eq:a}) and keeping only linear terms with respect to $\delta H$, we find:
\begin{equation}\label{eq:diffurforH}
\delta \ddot{H}+3H_0\delta \dot{H}+H_{0}^2\delta H\left(\frac{f_{,R}(H_0)}{48H_{0}^6\Phi_{,\mathcal{G}\mathcal{G}}(H_0)+3H_{0}^2f_{,RR}(H_0)}-4\right)=0
\end{equation}
We are seeking solutions in the form $\delta H=C_1e^{\lambda_{+}t}+C_2e^{\lambda_{-}t}$, and substituting it into Eq.(\ref{eq:diffurforH}), we obtain an algebraic equation with the following roots:
\begin{equation}
\lambda_{\pm}=\frac{3H_{0}}{2}\left(-1\pm \sqrt{1-\frac{4}{9}\left(\frac{f_{,R}(H_0)}{48H_{0}^6\Phi_{,\mathcal{G}\mathcal{G}}(H_0)+3H_{0}^2f_{,RR}(H_0)}-4\right)} \right)
\end{equation}

This shows that the de Sitter point is stable under the condition:
\begin{equation}
\frac{48H_{0}^6\Phi_{,\mathcal{G}\mathcal{G}}(H_0)+3H_{0}^2f_{,RR}(H_0)}{f_{,R}(H_0)}<\frac{1}{4}
\end{equation}
in agreement with \cite{Ali}.
For $f(R)=R+\alpha R^n$, $\Phi(\mathcal{G})=\beta \mathcal{G}^{\frac{n}{2}}$ we easily get:
\begin{equation}
	H_{0}^{2n-2}\frac{\beta n(n-2)2^{\frac{n}{2}}12^{\frac{n}{2}-1} +\alpha n(n-1)12^{n-1}}{1+\alpha n12^{n-1} H_{0}^{2n-2}}<1\,.
		\end{equation}
Substituting $H_0$ from the equation of motion (\ref{eq:dSRRnGm}), we get the following stability condition of the de Sitter solution:
\begin{align}
&\frac{\beta}{\alpha} <-(6^{\frac{n}{2}}), \quad n>2\,,\\
&\frac{\beta}{\alpha} >-(6^{\frac{n}{2}}), \quad 1<n<2\,,\\
&\frac{\beta}{\alpha} <-(6^{\frac{n}{2}}), \quad n<1\,.
		\end{align}
		\par 
Comparing this conditions with the stability situation for power-law solutions in $R^n+\beta \mathcal{G}^{\frac{n}{2}}$ we can make a conclusion that there is some non-trivial connection between these solutions. 
When the de Sitter point becomes stable, stable power-law solution become non-phantom. When this power-law solution loses its stability, the de Sitter solution remains stable.

\section{Conclusions}
In this paper, we have considered the cosmological dynamics in modified gravity with both $f(R)$ and the Gauss-Bonnet term in the Lagrangian. In this theory, the case of $R^n+\beta\mathcal{G}^\frac{n}{2}$ appears to be special one. In this case the terms in the equations of motion, originating from these two different modifications of the action, are equally important for power-law solutions. The Dynamical system approach works well only in this case and helps us to find exact solutions and study their stability.  It was obtained that in the presence of Gauss-Bonnet term the general picture differs from that found for $f(R)$ and $f(\mathcal{G})$-gravity. There is a range of coupling constants in which there are stable non-phantom solutions for any integer value of $n>2$ as well as for $n=1$.

It has been shown that there are some difficulties when the dynamical system approach is applied to $f(R)+\Phi(\mathcal{G})$-gravity. For
power-law $f(R)$ and $\Phi(\mathcal{G})$ studied in this paper 
 it works well only while both terms are equally important on power-law solutions for the scale factor.
It is useless for transient regimes, where only one term dominates. Thus we cannot investigate the Universe dynamics using only this method because different terms can dominate at different stages of its evolution. It should also be noted that some cosmological solution can be lost
when transforming the initial system to a set of first-order differential equations in expansion normalized variables, and the choice of these variables
may be sometimes tricky (in our paper we use two different system of variables to consider $n=1$ and $n>1$ cases).  
  
We have also studied in detail the existence and stability conditions of a de Sitter solution for general power-law functions $f(R)$ and $\Phi(\mathcal{G})$.

\section*{Acknowledgments}
This work was supported by Federal
Russian Science Agency through the research contract
02.740.11.0575 and RFBR grant 11-02-00643

\end{document}